\documentclass[a4paper]{article}
\usepackage{ASVspoof}
\usepackage{epsfig,amssymb,amsmath}
\usepackage{cite}
\usepackage{multirow}
\usepackage{graphicx}
\usepackage{booktabs}
\usepackage{amssymb}
\usepackage{caption}
\usepackage[bottom]{footmisc}

\ninept

\title{XMUspeech Systems for the ASVspoof 5 Challenge}

\name{}

\makeatletter
\def\name#1{\gdef\@name{#1\\}}
\makeatother
\name{{\em Wangjie Li\textsuperscript{1}, Xingjia Xie\textsuperscript{1}, Yishuang Li\textsuperscript{1,2},}\\
{\em Wenhao Guan\textsuperscript{3}, Kaidi Wang\textsuperscript{3}, Pengyu Ren\textsuperscript{3}, Lin Li\textsuperscript{*1,2}, Qingyang Hong\textsuperscript{*3}\thanks{*Corresponding authors}}}

\address{\textsuperscript{1}School of Electronic Science and Engineering, Xiamen University, China\\
\textsuperscript{2}Institute of Artificial Intelligence, Xiamen University, China\\
\textsuperscript{3}School of Informatics, Xiamen University, China\\
{\small \tt liwangjie@stu.xmu.edu.cn,\{lilin,qyhong\}@xmu.edu.cn} }
\begin{document}
\maketitle

\begin{abstract}
In this paper, we present our submitted XMUspeech systems to the speech deepfake detection track of the ASVspoof 5 Challenge. Compared to previous challenges, the audio duration in ASVspoof 5 database has significantly increased. And we observed that merely adjusting the input audio length can substantially improve system performance. To capture artifacts at multiple levels, we explored the performance of AASIST, HM-Conformer, Hubert, and Wav2vec2 with various input features and loss functions. Specifically, in order to obtain artifact-related information, we trained self-supervised models on the dataset containing spoofing utterances as the feature extractors. And we applied an adaptive multi-scale feature fusion (AMFF) method to integrate features from multiple Transformer layers with the hand-crafted feature to enhance the detection capability. In addition, we conducted extensive experiments on one-class loss functions and provided optimized configurations to better align with the anti-spoofing task. Our fusion system achieved a minDCF of 0.4783 and an EER of 20.45\% in the closed condition, and a minDCF of 0.2245 and an EER of 9.36\% in the open condition.

\end{abstract}

\section{Introduction}
Speech serves as a crucial means of communication for human interaction, conveying not only the intended linguistic content but also the unique identity features of the speakers. In recent years, significant progress has been made in text-to-speech (TTS) and voice conversion (VC) technologies based on deep learning \cite{wang2021prosody,zhao2020voice,chen2024valle2neuralcodec}. These technologies can generate natural speech which is challenging for humans to distinguish from authentic speech. The malicious misuse of these deepfake technologies poses a serious threat to societal security and political stability.

To counteract the threat, it is imperative to progress in spoofing speech detection and keep it apace. Despite numerous attempts and commendable performance achieved in previous research on spoofing speech detection, the detection systems still exhibit poor generalization and robustness \cite{yi2023audio}. While many works have addressed these issues through loss functions \cite{zhang2021one,ding2023samo} and continual learning \cite{zhang2023you,zhang2024remember}, existing spoofing speech detection methods are prone to overfitting on the training set. And they usually lack robustness in detecting processed speech through heavy compression, noise addition, etc. 

The automatic speaker verification spoofing and countermeasures (ASVspoof) challenge series is a community-led initiative that aims to promote the consideration of spoofing and speech deepfakes in addition to developing countermeasures. The ASVspoof 5 Challenge aims to benchmark the latest detection solutions in the face of more diverse and adversarial attacks \cite{Wang2024_ASVspoof5}.

In this paper, we describe our systems developed for the ASVspoof 5 Challenge, in particular for Track 1 in closed and open conditions. We explore the performance of state-of-the-art neural network architectures with different input features and loss functions. The rest of the paper is organized as follows: Section 2 describes the models and we applied for both conditions in Track 1. Section 3 presents the details about the datasets that we used and other experimental configurations. Results and analysis are presented in Section 4. Finally, we draw the conclusion in Section 5. 
\begin{figure}[t]
    \includegraphics[width=\columnwidth]{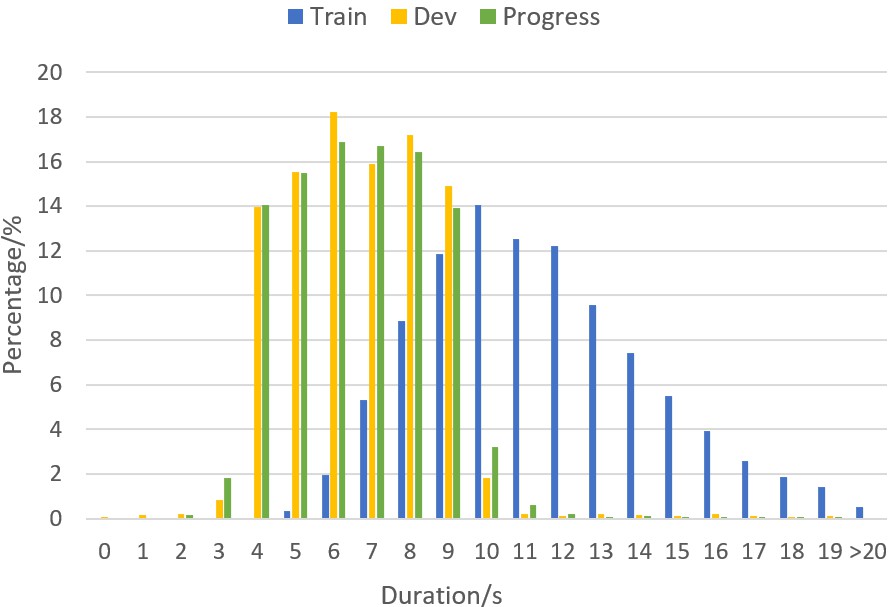}
    \caption{{\it Duration statistics of utterances in the training, development and progress datasets of ASVspoof 5 Challenge.}}
    \label{dur}
\end{figure}

\section{System Description}
\subsection{Data Augmentation}
The ASVspoof 5 Challenge maintains the pursuit of generalized countermeasures like previous challenges, which means we need to evaluate the threat of spoofing attacks with nonstudio-quality data which may be compressed using conventional codecs or contemporary neural codecs \cite{delgado2024asvspoof}. Referring to the configuration of ASVspoof 2021, we applied a random codec offline to each training utterance of ASVspoof 5 Challenge, amplifying twice the amount of training data. As in previous study \cite{cohen2022study,ferras2016large}, the codec was chosen from different compression methods with variable bitrates and channel effect simulators with bandwidth limits. However, this augmentation did not lead to any performance improvement. We suspect this is because the ASVspoof 5 Challenge training data already includes certain types of codecs. Adding more random codecs might not accurately reflect real-world scenarios. Such inappropriate data augmentation makes it difficult to improve performance. We also explored other augmentation techniques such as RawBoost \cite{tak2022rawboost} which includes convolutive and additive noise, SpecAugment \cite{park2019specaugment}, SpecAverage \cite{cohen2022study} and their combinations. Unfortunately, these augmentation techniques did not yield substantial improvements in our systems for the ASVspoof 5 Challenge.
 
\subsection{Model Architecture}
The artifacts of spoofing utterance may exist at spectral and temporal domains locally and globally \cite{jung2022aasist,liu2023leveraging,shin2024hm}. Integrating the dependency at different levels would therefore benefit the anti-spoofing systems. Thus, convolutional neural network (CNN), graph attention network (GAT), or Transformer are commonly used for the anti-spoofing countermeasure.

\subsubsection{AASIST}
GAT-based models leverage graph nodes to aggregate local dependency in the feature map. And they employ self-attention mechanism between these nodes to capture global artifacts across spectral and temporal domains \cite{tak2021graph,tak2021end,jung2022aasist,zhang2024improving}. Among these, AASIST \cite{jung2022aasist} exhibits superior performance by heterogeneous stacking graph attention layer and stack nodes, cooperating with a new max graph operation and an extended readout scheme. It is also applied as one of the ASVspoof 5 Challenge baselines.

\subsubsection{HM-Conformer}
Transformer captures global dependency efficiently through the self-attention mechanism \cite{vaswani2017attention}. For anti-spoofing, Rawformer \cite{liu2023leveraging} was proposed to capture local-global dependency by replacing GAT with Transformer in the AASIST, operating on entire sequence lengths. 
Furthermore, Conformer architecture \cite{gulati2020conformer} combines the CNN for extracting local features and the Transformer for exploring global dependency. And unlike the previous pipeline detectors, Conformer explores local-global features at the same time by fusing the Transformer encoder with the convolution module, considering the entire audio information. Shin et al. \cite{shin2024hm} proposed HM-Conformer by modifying the Conformer structure through hierarchical pooling and multi-level classification token aggregation (MCA) method, which achieved outperforming results on the ASVspoof 2021 Deepfake dataset. The hierarchical pooling conveys more compact features to reduce duplicated information and the MCA method aggregates task-related information from various encoder blocks. 

\subsubsection{Hubert}
Recently, a series of Transformer-based self-supervised models (SSMs) have been applied successively in the anti-spoofing \cite{tak2022automatic,guo2024audio,kawa2023improved,yang2024robust}. SSMs can learn effective universal speech representations from utterances. Compared to traditional acoustic features, SSMs extract more valuable information and contribute to improving the performance of downstream tasks \cite{sankala2022multi,yang2021superb}. Although many studies have demonstrated the powerful performance of SSMs with supervised fine-tuning, these models did not take anti-spoofing as the training purpose initially. They performed unsupervised clustering only on bonafide speech to extract features. We would like to further explore the performance of a self-supervised model trained on a dataset that contains bonafide and spoofing data. We chose Hubert \cite{hsu2021hubert} architecture as the backbone since it performed excellent generalization \cite{yang2024robust}. We trained a Hubert of base size as a feature extractor using ASVspoof 5 Challenge training data, with AASIST or HM-Conformer as a classifier.

Due to the fact that different Transformer layers in the SSMs can learn speech representations at various hierarchies \cite{fan2020exploring,chen2022does}, our previous work \cite{li2024sr} has proposed an adaptive multi-scale feature fusion module (AMFF) to model local dependency that distributes across different Transformer layers. The AMFF is sequentially stacked with an average pooling layer, a FC layer, a ReLU activation function, another FC layer, and a Sigmoid activation function, aiming to acquire multi-scale speech representations with artifact-related information. And we applied AMFF to integrate features from multiple Transformer layers with the hand-crafted feature to enhance the detection capability.

\subsection{Loss Function}
The ASVspoof 5 Challenge remains on the generalization ability to unseen attacks. One-class classification approach is an appealing method to address the problem. Zhang et al. \cite{zhang2021one} proposed OC-Softmax loss function as the training objective to compact bonafide speech into one cluster in the learned embedding space while pushing away spoofing attacks from the cluster. This approach has shown significant improvement in the generalization ability to unseen spoofing attacks. 

However, due to the variety of timbre and speaking traits of different speakers, Zhang et al. \cite{ding2023samo} have found that the bonafide speech of different speakers naturally forms multiple clusters in the embedding space. Only one cluster to compact different speakers may have caused the misclassification of the spoofing attacks. Based on OC-Softmax loss, they proposed speaker attractor multi-center one-class learning (SAMO), which clusters bonafide speech around a number of speaker attractors and pushes away spoofing attacks from all the attractors in a high-dimensional embedding space.
These clusters are formed based on speaker identity during training, which are called attractors. During inference, the cosine similarity between the test utterance and the nearest or corresponding speaker attractor (depending on whether there is enrollment data) is to score each trial as bonafide or spoof.

\begin{table*}[t]
    \caption{Comparison of backbone models with different configurations. The configuration \em update interval \em =1 means updating SAMO loss attractors every epoch. We apply \em weighted loss \em according to category proportion. The \em maxscore \em means calculating the similarity between every utterance and its nearest speaker attractor for SAMO loss. The \em enrollment \em indicates calculating similarity with enrollment utterances during inference.}
    \label{model compare}
    \resizebox{\textwidth}{!}{%
    \begin{tabular}{@{}ccccccccccc@{}}
    \toprule
    \multirow{2}{*}{\textbf{ID}} & \multirow{2}{*}{\textbf{Backbone}} & \multirow{2}{*}{\textbf{Loss}} & \multicolumn{4}{c}{\textbf{Configuration}} & \multicolumn{2}{c}{\textbf{Dev}} & \multicolumn{2}{c}{\textbf{Progress}} \\ \cmidrule(l){4-11} 
     &  &  & update interval & weighted loss & maxscore & enrollment & minDCF & EER(\%) & minDCF & EER(\%) \\ \midrule
    B1 & AASIST & Softmax & - & \checkmark & - & - & 0.32 & 15.20 & \textbf{0.42} & \textbf{16.32} \\
    S2 & \multirow{2}{*}{HM-Conformer} & OC-Softmax & - & × & - & - & 0.40 & 16.96 & 0.75 & 25.83 \\
    S3 &  & OC-Softmax & - & \checkmark & - & - & 0.29 & 14.38 & 0.64 & 24.88 \\
    S4 & \multirow{8}{*}{HM-Conformer} & SAMO & 1 & × & × & × & 0.31 & 15.60 & 0.47 & 18.12 \\
    S5 &  & SAMO & 1 & × & × & \checkmark & 0.29 & 15.47 & 0.44 & 17.07 \\
    S6 &  & SAMO & 3 & × & × & \checkmark & 0.37 & 16.26 & 0.48 & 21.64 \\
    S7 &  & SAMO & 5 & × & × & \checkmark & 0.38 & 24.82 & 0.56 & 27.44 \\
    S8 &  & SAMO & 1 & × & \checkmark & \checkmark & 0.30 & 15.22 & 0.45 & 17.82 \\
    S9 &  & SAMO & 3 & × & \checkmark & \checkmark & 0.30 & 15.35 & 0.46 & 18.17 \\
    S10 &  & SAMO & 1 & \checkmark & \checkmark & \checkmark & 0.28 & 15.61 & 0.42 & 17.56 \\
    S11 &  & SAMO & 3 & \checkmark & \checkmark & \checkmark & \textbf{0.28} & \textbf{12.32} & 0.48 & 21.56 \\ \bottomrule
    \end{tabular}%
    }
\end{table*}

\begin{table*}[h]
    \caption{Details and results of XMUspeech systems submitted for Track 1 in the closed and open condition.}
    \label{detail models}
    \resizebox{\textwidth}{!}{%
    \begin{tabular}{@{}ccccccccccc@{}}
    \toprule
    \multirow{2}{*}{\textbf{ID}} & \multirow{2}{*}{\textbf{Dataset}} & \multirow{2}{*}{\textbf{Feature}} & \multirow{2}{*}{\textbf{Backbone}} & \multirow{2}{*}{\textbf{Loss}} & \multicolumn{2}{c}{\textbf{Dev}} & \multicolumn{2}{c}{\textbf{Progress}} & \multicolumn{2}{c}{\textbf{Eval}} \\ \cmidrule(l){6-11} 
     &  &  &  &  & minDCF & EER(\%) & minDCF & EER(\%) & minDCF & EER(\%) \\ \midrule
    B1 & ASVspoof 5 & - & AASIST & Softmax & 0.31 & 15.20 & 0.42 & 16.32 & 0.71 & 29.12 \\
    C2 & ASVspoof 5 & - & AASIST & Softmax & 0.28 & 15.16 & 0.32 & 12.28 & \multicolumn{2}{c}{\multirow{5}{*}{-}} \\
    C3 & ASVspoof 5 & LFCC & HM-Conformer & SAMO & 0.28 & 15.61 & 0.42 & 17.56 & \multicolumn{2}{c}{} \\
    C4 & ASVspoof 5 & Hubert & AASIST & Softmax & 0.12 & 5.23 & 0.33 & 16.60 & \multicolumn{2}{c}{} \\
    C5 & ASVspoof 5 & Hubert & HM-Conformer & SAMO & 0.13 & 5.41 & 0.39 & 20.33 & \multicolumn{2}{c}{} \\
    C6 & ASVspoof 5 & Hubert+LFCC & HM-Conformer & SAMO & 0.27 & 13.09 & 0.38 & 14.66 & \multicolumn{2}{c}{} \\
    Fusion(B1-C6) & - & - & - & - & 0.12 & 5.01 & 0.19 & 8.01 & 0.48 & 20.45 \\ \midrule
    O7 & Librispeech & Wav2vec & AASIST & Softmax & 0.14 & 5.66 & 0.28 & 12.33 & \multicolumn{2}{c}{\multirow{3}{*}{-}} \\
    O8 & Librispeech & Wav2vec & HM-Conformer & SAMO & 0.17 & 7.97 & 0.36 & 16.34 & \multicolumn{2}{c}{} \\
    O9 & Librispeech & Wav2vec+LFCC & AASIST & Softmax & 0.16 & 7.50 & 0.30 & 13.44 & \multicolumn{2}{c}{} \\
    Fusion(B1-O9) & - & - & - & - & 0.09 & 3.94 & 0.11 & 4.97 & 0.23 & 9.36 \\ \bottomrule
    \end{tabular}%
    }
\end{table*}

\begin{figure}[t]
    \includegraphics[width=\columnwidth]{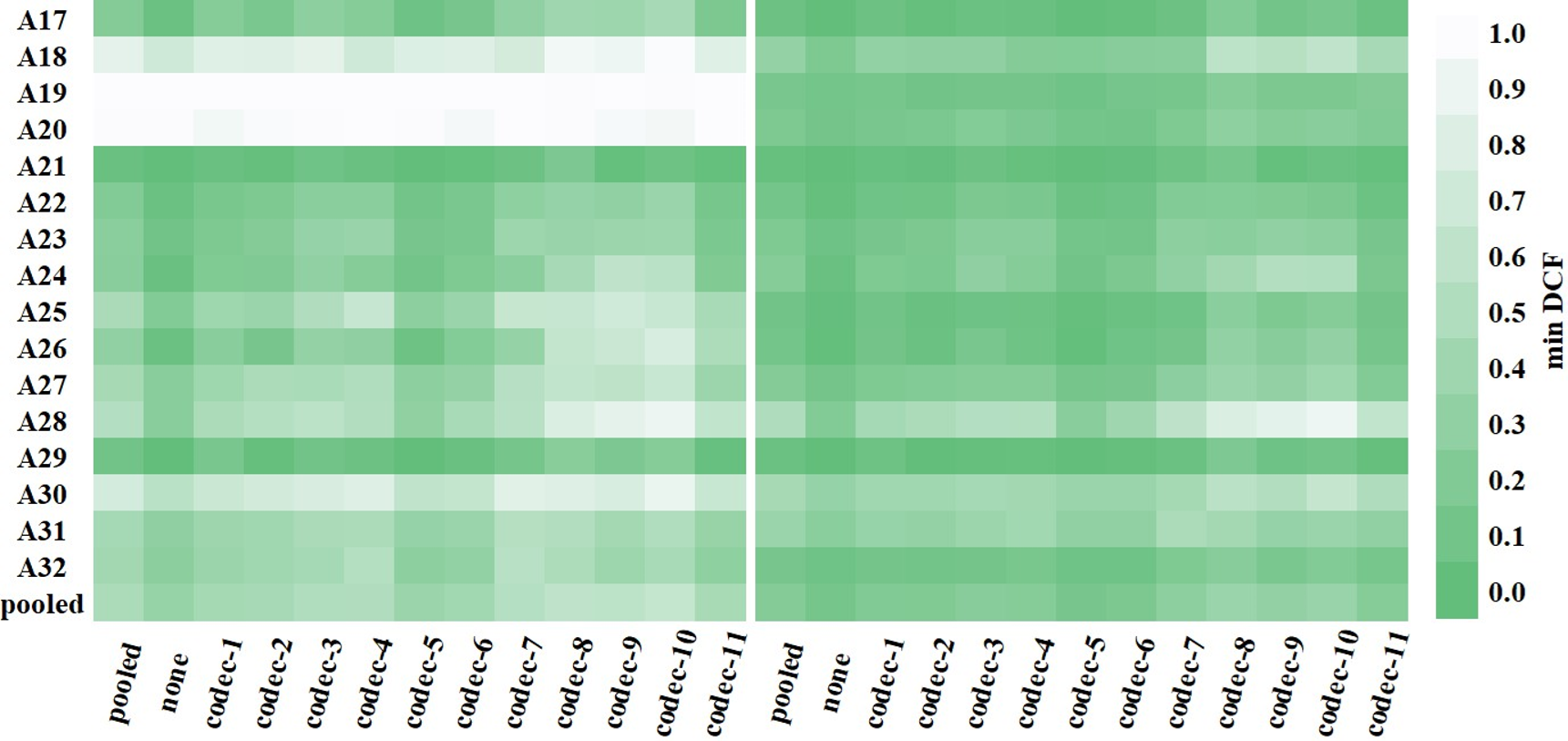}
    \caption{{\it Detailed result breakdown of our submitted systems in closed condition (left) and open condition (right). A17-A32 represent diverse spoofing attacks and codec 1-11 are different codecs and compression conditions in the evaluation dataset. The darker the color, the better the performance.}}
    \label{breakdown}
\end{figure}

\section{Experimental Setup}
\subsection{Dataset}
The ASVspoof 5 Challenge database is built from crowdsourced data collected from a vastly greater number of speakers in diverse acoustic conditions \cite{Wang2024_ASVspoof5}. Different from the previous ASVspoof challenges with most speech in length of 3 s to 5 s, the average duration of the ASVspoof 5 Challenge database is up to 10 s as shown in Figure \ref{dur}. This may result in systems trained with the speech of 4 s performing worse during the ASVspoof 5 Challenge evaluation, since long utterances tend to provide more information making it easier to capture artifacts \cite{zhang2024improving,gulati2020conformer}.

The ASVspoof 5 Challenge set up a new open condition that allows participants to use external data and pre-trained speech foundation models, subject to there being no overlap between training data and evaluation data \cite{Wang2024_ASVspoof5}. For the open condition, we used Librispeech \cite{panayotov2015librispeech} and corresponding pre-trained speech foundation models of base size. 

Additionally, we employed speech synthesis techniques to generate more spoofing data for the open condition. Here we utilize the LibriTTS \cite{zen2019libritts} and all the bonafide utterances in ASVspoof 5 Challenge training data to generate about 500 h extra spoofing data. The configurations include three type models: HiFi-GAN \cite{kong2020hifi}, BigVGAN \cite{lee2022bigvgan}, and our previous work ReflowTTS \cite{guan2024reflow}. The data generation can be categorized into two approaches: directly reconstructing audio using a vocoder and generating audio through an acoustic model combined with a vocoder.
\begin{itemize}
\item Vocoder Reconstruction: We use two vocoders, HiFi-GAN and BigVGAN, for reconstruction. After training the models on LibriTTS, we directly extract the mel spectrograms of the bonafide utterances and use both vocoders to generate the corresponding spoofing utterances.
\item Acoustic Model + Vocoder: Our acoustic model is based on the ReflowTTS architecture, with an added speaker encoder to provide speaker information. During training, we randomly crop 3-second segments of the current utterance's mel spectrograms as the input for the speaker encoder. After training, we randomly select a text from the LibriTTS dataset and combine it with speaker information from a bonafide utterance. We generate the corresponding mel spectrograms with the text content and speaker information, and then generate spoofing utterances using both HiFi-GAN and BigVGAN.
\end{itemize}
\subsection{Feature Extraction}
Following previous works \cite{jung2022aasist,shin2024hm}, we randomly truncated or repeated the audio clips to 10 s as the input chunk size for each batch during the training stage. For inference, we truncated the first 10 s of the utterance or repeated the audio to 10 s, keeping the consistency of every inference. Since the audio input length changed from 4 s to 10 s, the stride of the first convolutional model in AASIST was set to 2, reducing the number of temporal nodes to 36. It should be noted that the audio input length of all systems was set to 10 s in subsequent experiments, except for the baseline.

For input features in HM-Conformer, we used 1000 frames of the 120-dimensional linear frequency cepstral coefficients (LFCC), encompassing a window length of 20 ms, a hop size of 10 ms, a 512-point FFT, a linearly spaced triangle filter bank of 40 channels, and delta and delta-delta coefficients. 

Additionally, we utilized the AMFF to fuse the multi-layer representations and also to integrate the fused representation with the hand-crafted feature, which is similar to our previous work \cite{li2024efficient}.

\subsection{Ensemble Systems}
Ensemble systems where each subsystem is tuned to some specific artifacts usually obtain excellent performance and reliable detection. We perform the logistic regression with Bosaris \cite{brummer2013bosaris} toolkit on development and progress datasets to fuse several sub-systems as shown in Table \ref{detail models}.

\subsection{Evaluation Metrics}
The spoofing countermeasure systems (CMs) need to assign a real-valued bonafide-spoof detection score to each utterance. Different from past ASVspoof challenge editions for which the Equal Error Rate (EER) was used as the primary metric for the comparison of spoofing CMs, ASVspoof 5 Challenge builds upon a normalized Detection Cost Function (DCF) \cite{sadjadi2020nist} for the Track 1. Details can be found in ASVspoof 5 Challenge summary paper \cite{Wang2024_ASVspoof5}. We take the minimum detection cost function (minDCF) and EER as the main metrics in the following analysis. 

\subsection{Training Details}
We employed the softmax loss function with a loss weight of 0.9 for bonafide utterances and 0.1 for spoof ones. For OC-Softmax loss function, we set  hyperparameters $\alpha = 20$, $m_0 = 0.9$, and $m_1 = 0.2$. As for SAMO loss function, we took all bonafide utterances in the ASVspoof 5 Challenge training dataset to form 400 loss attractors with the margins $m_0 = 0.7$ and $m_1 = 0$ for training. And we used enrollment utterances for inference. In addition, we trained all systems on four NVIDIA A40 GPUs with a learning rate of $10^{-4}$ and cosine annealing learning rate decay for 100 epochs. We utilized Adam \cite{kingma2014adam} optimizer with $\beta_1 = 0.9$ and $\beta_2 = 0.999$ and a batch size of 300. Then we saved the model that achieved the lowest minDCF on the development set.

\section{Results and Analysis}

\subsection{Model Comparison}
Table \ref{model compare} compares two backbone models with different loss functions and configurations in the progress phase. We mainly explored the performance of HM-Conformer combined with different loss functions. We found that applying weighted loss according to the category ratio of bonafide and spoof utterances can improve system performance. 

For SAMO configuration, using enrollment utterances of corresponding datasets to form the loss attractors during inference can bring certain performance improvement, which is consistent with \cite{ding2023samo}. We also found that when calculating the cosine similarity in the training stage, taking the distance between every utterance and its nearest loss attractor as the loss value instead of corresponding loss attractor by speaker identity, will achieve better performance. Considering there are no speaker labels during inference, we could only take the maxscore of cosine similarity between test utterance and SAMO loss attractors as the output score. Therefore, it is reasonable to also apply maxscore during the training stage to keep training and inference consistent. And we believe it may be too difficult for an anti-spoofing system to match the utterance with the corresponding loss attractor by speaker identity. 
By calculating the distance to the nearest loss attractor, we weaken the speaker attribute of the loss attractors and focus on their bonafide characteristics, so that the system can adapt better to the anti-spoofing task.

\begin{figure}[t]
 \includegraphics[width=\columnwidth]{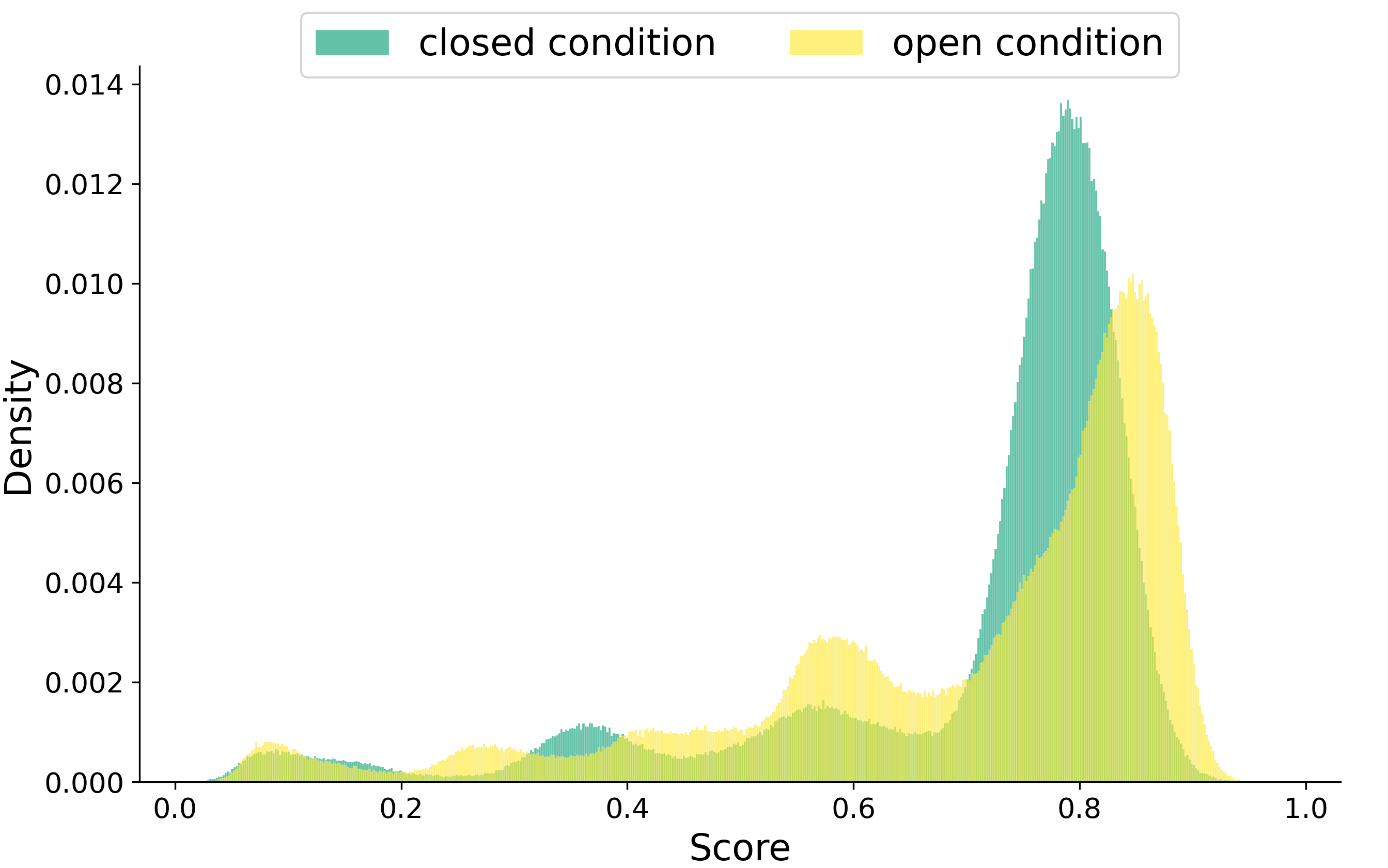}
 \caption{{\it Score density histogram of the submitted system for closed and open condition.}}
 \label{score}
\end{figure}
\subsection{Performance Analysis of Submitted Systems}

The overall performance of our submitted systems in the progress and evaluation phase is shown in Table \ref{detail models}. In the closed condition, we are surprised to find that just by increasing the input audio length of the baseline AASIST to the average duration of the ASVspoof 5 Challenge datasets, the performance was significantly optimized, with an improvement of 24\% relatively for minDCF on the progress dataset. While HM-Conformer using the hand-crafted feature as input also achieves competitive performance compared to the baseline.

The systems using Hubert which is trained on the ASVspoof 5 Challenge training data as the feature extractor, show excellent performance on the development set, especially compared with the original systems. But the use of SSMs also causes obvious overfitting when evaluated on the progress dataset. The system with fused features composed of Hubert and hand-crafted feature achieves consistent performance on the development and progress datasets without obvious overfitting. Compared with the baseline, it improves the performance by nearly 10\% for minDCF on the progress dataset. The final fusion system achieves results of minDCF 0.478 and EER 20.45\% on the evaluation set under the closed condition, improved by 33\% and 30\% respectively compared to the baseline.

For the open condition, we used Wav2vec2.0 \cite{baevski2020wav2vec} of the base size which is pre-trained on Librispeech and we fine-tuned it using the ASVspoof 5 Challenge training data and additional generated spoofing data mentioned in Section 3.1. It can be found that the overall performance of systems trained in the open condition performs better than the systems under the closed condition, but over-fitting also occurs here on the development and progress datasets. The final fusion system achieved results of minDCF 0.225 and EER 9.36\%, which are relatively 68\% and 67\% better than the baseline respectively.

Figure \ref{breakdown} reports the detailed result of our submitted system for the closed condition with different attacks and codecs on the evaluation dataset. Among diverse spoof attacks, we observe that A19 (MaryTTS) and A20 (an adversarial attack) are the most difficult attack types for our system to recognize. From the perspective of codecs, codec-10 (speex with bandwidth 8 kHz and low bitrates) is the hardest one to detect. The configuration details of attacks and codecs can be found in \cite{Wang2024_ASVspoof5}. As for the open condition shown in Figure \ref{breakdown}, A19 and A20 are no longer as threatening as in closed conditions, but the codec-10 remains tough for the system to detect.

The score distribution of our submitted system for the closed and open condition is presented in Figure \ref{score}. We can see that the distribution trend of scores is generally the same whether in closed or open condition, but the score distribution in the open condition is more discriminative, which also indicates that the system trained in the open condition achieves more reliable performance. 

\section{Conclusion}
In this paper, we introduced the XMUspeech systems for the speech deepfake detection task in both closed and open conditions. Our investigation focused on models like AASIST and HM-Conformer, which capture discriminative artifacts both locally and globally. We evaluated their performance with various input features and loss functions using the ASVspoof 5 Challenge database. A significant finding was that increasing the audio duration of input utterances substantially improves performance. 

We conducted extensive experiments to explore the performance of one-class learning loss functions, such as OC-Softmax and SAMO. We recommend strengthening the bonafide attributes of SAMO loss attractors and weakening their speaker identity information by calculating the cosine similarity distance to the nearest loss attractor during training. This allows the systems to better focus on the anti-spoofing task.

Furthermore, we explored the performance of self-supervised models trained on datasets containing spoofing utterances to extract effective artifact-related representations. We successfully enhanced detection capabilities by utilizing an adaptive multi-scale feature fusion model, integrating features from multiple Transformer layers and hand-crafted features. The use of diverse acoustic features greatly intensified the diversity and complementarity of our systems.

These approaches enabled our final fusion system to achieve significant improvements compared to the baseline. However, our results also indicate that adversarial attacks pose a significant threat to our systems. In the future work, we will further explore effective methods to defend against these adversarial attacks.

\bibliographystyle{IEEEbib}
\bibliography{ASVspoof_BibEntries}

\end{document}